\documentclass[runningheads]{llncs}

\usepackage[T1]{fontenc}
\usepackage[scale=0.9]{inconsolata}

\usepackage{algorithm}
\usepackage{algcompatible}
\usepackage{amsmath,amsfonts,amssymb}
\usepackage{graphicx}
\usepackage{booktabs}
\usepackage{cite}
\usepackage{float}
\usepackage{tabularx}
\usepackage{url}
\usepackage{xcolor}
\usepackage{listings}
\usepackage{multirow}
\usepackage{enumitem}
\usepackage{xspace}
\usepackage{breakurl}
\usepackage[breaklinks]{hyperref}

\setlist[itemize]{leftmargin=1.15em,labelsep=0.35em,topsep=2pt,itemsep=0pt,parsep=0pt}
\setlist[enumerate]{leftmargin=1.4em,labelsep=0.35em,topsep=2pt,itemsep=0pt,parsep=0pt}
\lstdefinestyle{python}{
  language=Python,
  basicstyle=\ttfamily\scriptsize,
  keywordstyle=\color{blue}\bfseries,
  stringstyle=\color{red},
  commentstyle=\color{gray}\itshape,
  showstringspaces=false,
  breaklines=true,
  frame=none
}

\newcommand{\code}[1]{{\texttt{#1}}}
\newcommand{\tool}{\textsc{ConcoLixir}\xspace}

\begin{document}

\title{\texorpdfstring{\textsc{ConcoLixir}}{ConcoLixir}: Reactive LLM Discovery Oracles for Python Concolic Testing\texorpdfstring{\thanks{This work was partially supported by the National Science and Technology Council (NSTC), Taiwan, under grant numbers 113-2221-E-004-010-MY2, 112-2222-E-004-001-MY3, and 114-2634-F-004-002-MBK. Chih-Duo Hong and Fang Yu are the corresponding authors.}}{}}
\titlerunning{\textsc{ConcoLixir}}

\author{Dong Chen \and Chih-Duo Hong \and Fang Yu}
\authorrunning{D. Chen et al.}
\institute{Department of Management Information Systems\\National Chengchi University\\Taipei, Taiwan}

\maketitle

\begin{abstract}
Concolic testing combines concrete execution with symbolic constraint solving, but Python programs expose recurring limits. Library calls can cause symbolic variables to downgrade to concrete values. Regular expressions, checksums, parsers, and other semantic operations can be hard to solve, and exploration can plateau on already covered paths.

We present \tool, a reactive LLM extension for Python concolic execution. The LLM acts as a discovery oracle, not a replacement for the solver or a correctness oracle. It generates initial seeds, proposes concrete inputs after solver failures, and targets uncovered code when coverage stalls. Each candidate is executed concolically, and only observed coverage and collected path constraints guide later exploration. Across synthetic, real-world, and library targets, \tool improves mean line coverage over the baseline concolic tester without an LLM oracle by 8.6, 15.1, and 17.0 percentage points. The gains are strongest near semantic barriers and library boundaries, and the full evaluation costs \$1.63 in API charges. These results show that bounded LLM discovery can complement symbolic reasoning without replacing it.

\end{abstract}

\section{Introduction}

\begin{figure}[t]
  \centering
  \includegraphics[width=.67\textwidth]{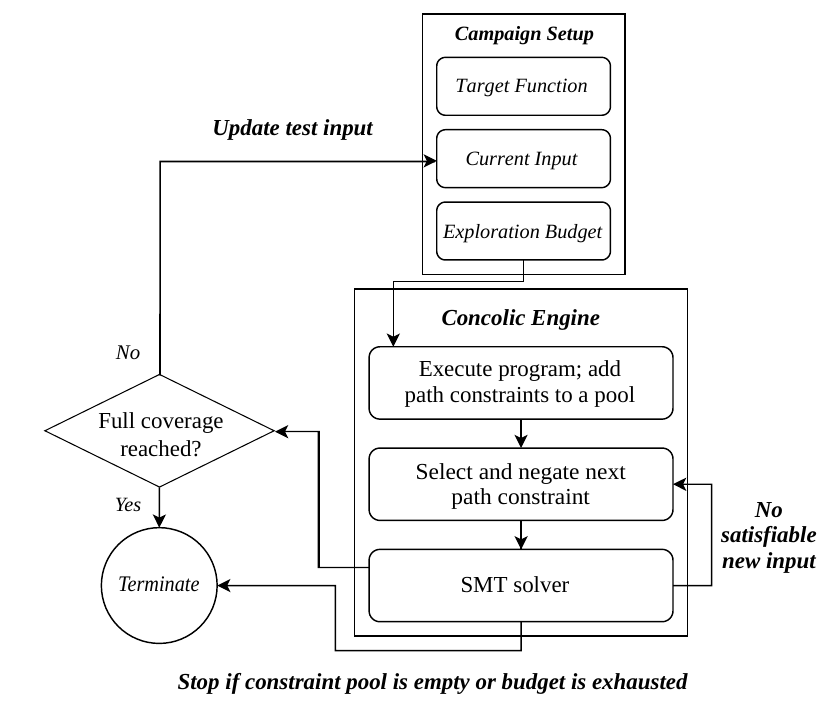}
  \caption{Baseline concolic testing loop driven by coverage. Starting from a current input, the engine executes the target and collects path constraints, selects and negates an unexplored path constraint, invokes the SMT solver to generate a new input when satisfiable, and repeats until full coverage is reached or the constraint pool or exploration budget is exhausted.}
  \label{fig:pyct_workflow}
\end{figure}

Concolic testing~\cite{sen2007concolic} generates tests by running a program on a concrete input, tracking symbolic path constraints at each branch, and asking a Satisfiability Modulo Theories (SMT) solver to create new inputs by negating selected constraints. This coverage-driven loop repeats until all reachable paths are explored or the budget runs out, as summarized in Figure~\ref{fig:pyct_workflow}.

Concolic testing has found defects in systems ranging from operating systems~\cite{cadar2008klee} to network protocols~\cite{godefroid2008sage}, and it is also used in modern test generation pipelines~\cite{stephens2016driller, yun2018qsym}. Python concolic tools, including \textsc{PyCT}, \textsc{CrossHair}, and \textsc{PyExZ3}~\cite{PyCT,crosshair,PyExZ3}, adapt this technique through \emph{object-oriented concolic substitution}~\cite{PyExZ3}, where each primitive argument is replaced by a proxy object that carries both a concrete value and a symbolic expression over the input variables. On Python programs from real projects, however, this approach faces three recurring challenges:

\emph{C1: Symbolic Downgrading.} Python programs often call external functions, including standard library routines and external packages, that may accept concolic arguments but return plain Python values. At these call boundaries, the symbolic expression is discarded and replaced by a concrete value, leaving later branches that depend on the return value without symbolic annotations for the SMT solver. For example, when a username validation function calls an unsupported string function, the operations run concretely inside the external function, and the symbolic expression for the string argument is lost.

\emph{C2: Constraints the Solver Cannot Handle Well.} Even when symbolic information is preserved, many paths in real code contain operations that the solver cannot handle well, leading to \textsc{Unknown} results, timeouts, or inaccurate models of library behavior. Modern solvers support integers, bitvectors, and basic string operations, but regex matching, JSON parsing, base64 decoding, and hash computations often fall outside that model, so the corresponding path remains unexplored.

\emph{C3: Coverage Plateaus.} Even when the solver succeeds, its new inputs may follow paths whose branches are already covered, so the constraint pool keeps shrinking without coverage growth and exploration stalls before reaching the iteration budget.

\begin{figure}[t]
  \centering
  \includegraphics[width=\textwidth]{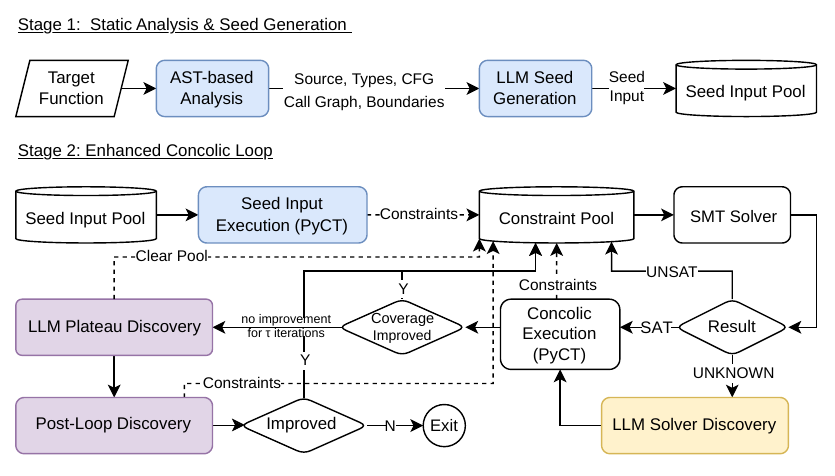}
  \caption{%
    \tool exploration flow. The SMT solver is consulted first. \tool uses three LLM roles (shaded boxes): seed generation, discovery after solver failure, and plateau discovery. The plateau role has two dispatch sites, one in the loop and one after the loop.}
  \label{fig:arch}
\end{figure}

These obstacles compound because symbolic downgrading removes branch predicates from the solver, while unsupported operations prevent the remaining predicates from being solved. Together, they make stale iterations more likely, since the search can keep revisiting already covered paths. Prior work addresses these problems separately. \textsc{PyCT}~\cite{PyCT} reduces downgrading through runtime function profiling but does not help when the solver times out or coverage stalls. \textsc{Driller}~\cite{stephens2016driller} and \textsc{QSYM}~\cite{yun2018qsym} break plateaus through mutation but target binary programs. LLM test generation~\cite{deng2023largelanguagemodelszeroshot, lemieux2023codamosa} treats the LLM as a standalone generator rather than integrating it into the concolic loop.

This combination raises whether an LLM can be integrated into the inner loop of a concolic engine to mitigate recurring instances of all three challenges without replacing symbolic execution. The answer is not obvious because LLMs are not guaranteed to be correct, their outputs vary across calls, and API latency adds time to an already time critical loop. If the LLM produces unhelpful inputs, it consumes budget that the solver could have used.

The key insight is that the LLM can be useful without solving symbolic constraints, as long as it discovers concrete inputs that cross semantic barriers such as regular expressions, parsers, checksum checks, and opaque library calls. Once such an input reaches new code, ordinary concolic execution can resume systematic path exploration. Figure~\ref{fig:arch} shows how this discovery role is integrated into the concolic loop. We therefore call the LLM a \emph{discovery oracle}, using oracle in the narrow sense of a generator of candidate concrete inputs rather than a trusted solver for path constraints. The LLM does not decide path feasibility, program correctness, or coverage. Those decisions remain with execution, coverage measurement, and the solver, and only candidates that improve coverage or add useful constraints affect later exploration.

We present \tool, which uses a \emph{reactive} integration strategy that keeps the SMT solver as the main reasoning engine during exploration. Before exploration begins, the LLM generates seed inputs to populate the constraint pool and help cross opaque library boundaries the solver cannot observe (C1). During exploration, the LLM is called only after \textsc{Unknown}/\textsc{Error} solver results ({C2}) or coverage plateaus ({C3}). Each suggested input is executed concolically, producing new path constraints that the solver can use later. Plateau calls and calls after the loop stop as soon as they fail to improve coverage, and discovery after solver failure remains bounded by the global concolic iteration budget.

We evaluate \tool on 22 designed benchmarks, 22 library functions from real projects, and 75 automatically selected library entry points. Compared with \textsc{Concolic-Only}, \tool improves mean coverage by 8.6, 15.1, and 17.0 pp on these groups, with smaller gains where the solver already suffices and larger gains on code from real projects and on library code. A few regressions (B03, B15, B16 in Section~\ref{sec:rq1}) come from changes in exploration order caused by seeds under the fixed iteration budget, not from solver unsoundness. Across all target groups, the evaluation over 1{,}190 target runs costs \$1.63 in API charges with \code{gpt-4o-mini-2024-07-18}.

In summary, we make the following contributions:

\begin{enumerate}
\item We introduce \tool, a reactive architecture for Python concolic testing with LLM support, where the LLM proposes concrete candidate inputs while \textsc{PyCT} remains responsible for execution, validation, constraint collection, and exploration driven by the solver.
\item We design an oracle escalation hierarchy that invokes the LLM only for seeds before the loop, solver failures, and coverage plateaus. Plateau calls and calls after the loop stop after a round that does not improve coverage. Discovery after solver failure is bounded by the main loop, and mini exploration after the loop is bounded separately.
\item We evaluate the design on 22 designed benchmarks, 22 library functions from real projects, and 75 automatically selected library entry points from \code{sympy.ntheory} and {PyYAML}. The results show where the integration helps most: targets with parsers, regular expressions, checksums, structured inputs, and opaque library boundaries.
\end{enumerate}

The rest of the paper is organized as follows. Section~\ref{sec:related} reviews related work. Section~\ref{sec:design} presents the system design. Section~\ref{sec:eval} summarizes the evaluation, including the experimental architecture and results, and Section~\ref{sec:limitations} discusses limitations and concluding remarks.

\section{Related Work}
\label{sec:related}

\noindent\textbf{Symbolic and Concolic Execution.} Symbolic execution is a standard way to explore program paths systematically~\cite{cadar2013symbolic}. \textsc{DART}~\cite{godefroid2005dart} and \textsc{CUTE}~\cite{sen2005cute} introduced concolic testing. \textsc{EXE}~\cite{exe2006}, \textsc{KLEE}~\cite{cadar2008klee}, and \textsc{SAGE}~\cite{godefroid2008sage} scaled it to large C programs, and \textsc{Mayhem}~\cite{mayhem2018} showed its value at competition scale. Modern compiler-based engines such as \textsc{SymCC}~\cite{symcc2020} achieve high throughput by adding symbolic tracking directly to compiled code. SMT solvers including \textsc{Z3}~\cite{z3_2008} and \textsc{CVC5} are central to these systems. Python tools including \textsc{PyCT}, \textsc{CrossHair}, and \textsc{PyExZ3}~\cite{PyCT,crosshair,PyExZ3} adapt concolic testing through object-oriented substitution~\cite{PyExZ3}, but symbolic downgrading at external function boundaries still limits them.

\noindent\textbf{Model Robustness Testing and Verification.} A related line of work applies formal or concolic methods to properties of learned models rather than coverage of ordinary Python functions. Recent systems use concolic testing to search for individual fairness violations in neural networks~\cite{huang2025pyfair} and adversarial examples in Transformer classifiers~\cite{hong2025transformerrobustness}. Other robustness work verifies recurrent neural networks through abstraction refinement~\cite{lin2026rnnrobustness} or symbolic surrogate extraction~\cite{hong2025registerautomata}. These studies share the broader goal of combining concrete observations with symbolic reasoning, but their targets and properties differ from our setting: \tool explores Python function inputs to improve line coverage, not neural-network robustness specifications.

\noindent\textbf{Fuzzing and Search Based Testing.} \textsc{Driller}~\cite{stephens2016driller} and \textsc{QSYM}~\cite{yun2018qsym} combine concolic engines with greybox fuzzers such as \textsc{AFL++}~\cite{aflpp2020} for binary programs. Search based and property based tools such as \textsc{Pynguin}~\cite{pynguin2022}, \textsc{EvoSuite}~\cite{evosuite2011}, and \textsc{Hypothesis}~\cite{hypothesis2019} generate tests through genetic algorithms or random exploration. \textsc{Pynguin} also won the SBFT Python competition~\cite{sbft2024}. These tools generate complete test suites with assertions. They work at the test suite level rather than the single function input level. Their coverage metrics measure test suite adequacy, not path exploration completeness. For this reason, these tools are related systems rather than direct baselines for our single function concolic setting. The hybrid fuzzing tools (\textsc{Driller}, \textsc{QSYM}) target compiled code and share constraints between fuzzer and solver. They do not address the interpreted language challenges that arise in Python, such as symbolic downgrading and opaque library boundaries.

\noindent\textbf{LLM Test Generation.} \textsc{TitanFuzz}~\cite{deng2023largelanguagemodelszeroshot} uses LLMs to generate seed programs for fuzzing deep learning libraries. \textsc{CodaMOSA}~\cite{lemieux2023codamosa} interleaves LLM generated tests with search based generation when coverage plateaus, but the LLM and the search engine operate independently. Each produces test cases without sharing symbolic constraints or path information. \textsc{CoverUp}~\cite{coverup2025} drives LLM test generation with iterative coverage feedback. \textsc{LLM-Sym}~\cite{wang2024llmsym} translates Python path constraints into \textsc{Z3} code, extending symbolic execution to list operations. These approaches treat the LLM as a standalone generator or translator. The LLM does not participate in the concolic loop or share state with the solver.

\noindent\textbf{Concolic Execution with LLM Integration.} Recent work integrates LLMs into concolic testing in different ways. \textsc{ConcoLLMic}~\cite{luo2026concollmic} replaces the symbolic engine with LLM agents. One agent summarizes execution traces into natural language or source level path constraints. Another agent generates test inputs that satisfy those constraints. This design is agnostic to programming language because the LLM can reason about any language without a dedicated symbolic backend. However, it uses the LLM heavily. Each test input iteration calls multiple agents to summarize a path constraint and synthesize a satisfying input, leading to 48\,h per subject with claude-3.7-sonnet. \textsc{AutoBug}~\cite{li2025autobug} similarly replaces the solver by breaking path analysis into subtasks for the LLM.

\textsc{CottonTail}~\cite{tu2026cottontail} takes a proactive approach. Its LLM-driven constraint solver uses a \emph{Solve-Complete} paradigm to generate valid structured inputs for parsers (XML, SQL, JavaScript, JSON). It queries the LLM for every constraint and also for seed acquisition when coverage saturates. Both \textsc{ConcoLLMic} and \textsc{CottonTail} operate on compiled programs. In that setting, library functions are compiled into the analysis, so symbolic downgrading at external function boundaries does not arise. Their instrumentation and pipelines for extracting constraints target compiled binaries. A direct empirical comparison is outside the scope of this study because their published pipelines target compiled settings and would require a different front end for interpreted Python functions.

\begin{table}[t]
  \caption{Comparison with approaches that use LLMs in test generation.}
  \label{tab:related}
  \centering
  \scriptsize
  \setlength{\tabcolsep}{2pt}
  \renewcommand{\arraystretch}{1.05}
  \begin{tabularx}{\columnwidth}{@{}
      >{\raggedright\arraybackslash}p{3.50cm}
      >{\raggedright\arraybackslash}p{1.95cm}
      >{\raggedright\arraybackslash}p{1.65cm}
      >{\raggedright\arraybackslash}X
    @{}}
    \toprule
    \textbf{Tool} & \textbf{LLM role} & \textbf{Target} & \textbf{Invocation and control} \\
    \midrule
    \multicolumn{4}{@{}l}{\textit{LLM as primary test generator}} \\
    \textsc{TitanFuzz}~\cite{deng2023largelanguagemodelszeroshot} & seed gen. & Python/DL & pre-run; primary generation \\
    \textsc{CodaMOSA}~\cite{lemieux2023codamosa} & test gen. & Python & plateau; primary generation \\
    \textsc{CoverUp}~\cite{coverup2025} & test gen. & Python & coverage feedback; primary generation \\
    \midrule
    \multicolumn{4}{@{}l}{\textit{LLM replaces or translates for the solver}} \\
    \textsc{LLM-Sym}~\cite{wang2024llmsym} & translator & Python & per translation; no coverage gate \\
    \textsc{ConcoLLMic}~\cite{luo2026concollmic} & agent loop & multi-lang. & per iteration; no coverage gate \\
    \textsc{CottonTail}~\cite{tu2026cottontail} & prot. solver & binary & per constraint; no coverage gate \\
    \midrule
    \multicolumn{4}{@{}l}{\textit{LLM assists the solver (this work)}} \\
    \textbf{\tool} & \textbf{reactive oracle} & \textbf{Python} & \textbf{seed; solver failure; coverage-gated plateau} \\
    \bottomrule
  \end{tabularx}
  \par\smallskip
  \footnotesize
  ``gen.'' = generator. ``prot.'' = proactive. Calls gated by coverage stop after no improvement. Discovery after solver failure is limited by the iteration budget.
\end{table}

Table~\ref{tab:related} summarizes the comparison. \tool takes a different approach for Python. It does not ask the LLM to formulate or solve constraints. Instead, it uses the LLM to find inputs that cross code regions the SMT solver cannot reach, such as opaque library calls, constraints the solver cannot handle well, and coverage plateaus. \textsc{PyCT} then executes these inputs concolically. This execution walks the newly reachable paths and collects path constraints that the solver \emph{can} handle. The LLM does not replace the solver. It moves the engine past barriers so that the solver can resume useful exploration on the other side.

\section{System Design}
\label{sec:design}

Figure~\ref{fig:arch} shows the complete exploration flow. The shaded boxes mark three LLM roles: seed generation before the main loop, discovery after solver failure during the loop, and plateau discovery. The plateau role has two dispatch sites. It is triggered in the loop when coverage stalls and after the loop for any remaining gaps.

\subsection{System Overview}

\tool extends \textsc{PyCT}~\cite{PyCT}, a Python concolic testing engine that implements \emph{object-oriented concolic substitution}~\cite{PyExZ3}. Each primitive argument is replaced by a proxy object (e.g., \code{ConcolicInt}, \code{ConcolicStr}). The proxy carries both a concrete value and a symbolic expression. At every branch point, the proxy's \code{\_\_bool\_\_} method records the branch predicate as a path constraint. The engine keeps a constraint pool. It explores the pool in breadth first order by negating a constraint, calling the SMT solver (\textsc{CVC5}), and executing the resulting input.

\tool extends this engine through a plugin interface. The kernel sends each LLM role to the plugin as a named event. The plateau role is split into two dispatch sites: one trigger in the loop and one push after the loop. As shown in Figure~\ref{fig:arch}, the system has two stages. Stage~1 performs static analysis and one LLM seed generation call before exploration begins. Stage~2 runs the enhanced concolic loop. It routes solver failures and coverage plateaus to the LLM, then runs a plateau discovery push after the loop to close any remaining gaps. The key design principle is the \emph{oracle escalation hierarchy}. During exploration, the SMT solver is tried first. LLM discovery is used only when solver progress fails or coverage stalls. LLM calls for plateaus stop as soon as they fail to improve coverage.

\subsection{Stage 1: Static Analysis and Seed Generation}
\label{sec:stage1}

Before any execution begins, an AST based static analysis extracts three kinds of information from the target function and its reachable callees up to a configurable depth. The analysis results are formatted into structured text for inclusion in LLM prompts.

\begin{itemize}
\item \emph{Call graph.} The analysis resolves function calls within the project boundary. It collects each callee's source code, signature, and call depth. Calls that cannot be resolved, such as external libraries and builtins, are recorded but not followed.

\item \emph{Branch conditions and boundary values.} For every reachable function, the AST visitor extracts all \code{if}/\code{while}/\code{assert} conditions as source text. It also collects numeric and string literals from comparison operators. These boundary values, such as threshold constants and enum strings, are aggregated across the entire call graph.

\item \emph{Control flow graph.} A CFG is extracted with nodes for each statement type, including \code{if}, \code{while}, \code{for}, \code{return}, and assignments. Its edges are labeled \emph{true}, \emph{false}, or \emph{unconditional}. The CFG enumerates up to 50 acyclic execution paths. When coverage data is available, the CFG can be filtered to show only paths that reach specific uncovered lines.
\end{itemize}

Once the static analysis completes, the LLM generates seed inputs that target all reachable lines. The prompt uses five context elements from the static analysis output. These are the function's source code with line numbers, type context with parameter types and custom type definitions, the CFG with execution paths, the call graph with branch conditions and boundary values from reachable callees, and instructions to cover all branches and boundary values.


\begin{figure}[t]
\begin{lstlisting}[basicstyle=\ttfamily\scriptsize, frame=single, xleftmargin=2pt, xrightmargin=2pt]
# Task: Generate test inputs for maximum coverage.

## Source Code:
  1| def func(param1: str, param2: int):
  2|     if check(param1):   # branch A
  ...

## Type Context:
  - param1: str
  - param2: int (range: 0..100)

## Control Flow Graph:
  Path 1: entry -> line 2 (true) -> line 3
  Path 2: entry -> line 2 (false) -> line 5

## Call Graph (branch conditions):
  check(): contains "param1 in valid_set"
  Boundary values: 0, 50, 100, "admin"

## Output:
  Return a Python list of dicts:
  [{"param1": ..., "param2": ...}, ...]
\end{lstlisting}
\caption{Abstract structure of a seed generation prompt. Live prompts include the full source code, complete CFG, and extracted boundary values.}
\label{fig:seed_prompt}
\end{figure}

The engine assembles these five elements into a prompt and sends it to the LLM (Figure~\ref{fig:seed_prompt}). The prompt asks the model to return a Python list of dictionaries inside a fenced code block. The response is parsed with \code{ast.literal\_eval}. A restricted eval fallback handles literals such as \code{"a" * 5} that \code{ast.literal\_eval} rejects. Any unparseable response is discarded. Unlike random seed generation, the model receives the full call graph context and tries to construct a small covering set.

Algorithm~\ref{alg:concolixir-merged} summarizes the complete campaign. Static analysis feeds seed generation. LLM seeds populate the constraint pool. The main loop tries the solver first and calls the LLM only on solver failure or coverage plateau. A final phase after the loop targets any remaining uncovered code.

\begin{algorithm}[!t]
\footnotesize
\caption{\tool Exploration Campaign}
\label{alg:concolixir-merged}
\begin{algorithmic}[1]
\REQUIRE target function $f$, default argument map $a_0$, configuration $C$
\ENSURE final line coverage for $f$ after the bounded exploration campaign
\STATE \COMMENT{Pre-loop seed discovery uses static context to cross early barriers.}
\STATE $G \gets \textsc{StaticAnalysis}(f)$
\STATE $S \gets \textsc{LlmSeedGeneration}(f,\;G)$
\STATE $W \gets \emptyset$;\;
       $T \gets \{a_0\}$;\;
       $\mathit{stale} \gets 0$
\FOR{each $s \in S$}
  \STATE $\textsc{ConcolicExec}(f,\;s,\;W)$;\;
         $T \gets T \cup \{s\}$
\ENDFOR

\STATE \COMMENT{Main loop keeps the SMT solver first and escalates only on stalls.}
\FOR{$\mathit{iter}=1$ to $C.\mathit{maxIter}$}
  \STATE \textbf{if} $W=\emptyset$ or $\textsc{FullCoverage}(f)$
         \textbf{then break}

  \IF{$\mathit{stale} \geq C.\tau$}
    \STATE $W \gets \emptyset$;\;
           $M \gets \textsc{UncoveredLines}(f)$
    \STATE $X \gets \textsc{LlmPlateau}(f,\;M,\;T)$
    \STATE \textbf{if} $\neg\textsc{TryCandidates}(X,\;f,\;W,\;T)$
           \textbf{then break}
    \STATE $\mathit{stale} \gets 0$;\; \textbf{continue}
  \ENDIF

  \STATE $\phi \gets \textsc{Pop}(W)$;\;
         $(\sigma,\rho) \gets \textsc{SmtSolve}(\phi)$

  \IF{$\rho \in \{\textsc{Unknown},\textsc{Error}\}$}
    \STATE $X \gets \textsc{LlmSolverDiscovery}
           (f,\;\textsc{UncoveredLines}(f),\;T)$
    \STATE \textbf{if} $\textsc{TryCandidates}(X,\;f,\;W,\;T)$
           \textbf{then} $\mathit{stale} \gets 0$
    \STATE \textbf{continue} \COMMENT{iteration-bound oracle}
  \ENDIF

  \STATE \textbf{if} $\rho=\textsc{Unsat}$ \textbf{then continue}
         \COMMENT{infeasible path}
  \STATE $a \gets \textsc{Merge}(a_0,\;\sigma)$;\;
         \textbf{if} $a \in T$ \textbf{then continue}
  \STATE $\mathit{cov} \gets \textsc{Coverage}(f)$;\;
         $\textsc{ConcolicExec}(f,\;a,\;W)$;\;
         $T \gets T \cup \{a\}$
  \STATE $\mathit{stale} \gets
         \textsc{UpdateStale}(f,\;\mathit{cov},\;\mathit{stale})$
\ENDFOR

\STATE \COMMENT{Post-loop discovery spends a small solver budget on remaining gaps.}
\FOR{$\mathit{round}=1$ to $C.r$}
  \STATE $M \gets \textsc{UncoveredLines}(f)$;\;
         \textbf{if} $M=\emptyset$ \textbf{then break}
  \STATE $X \gets \textsc{LlmPlateau}(f,\;M,\;T)$
         \COMMENT{post-loop discovery}
  \STATE \textbf{if} $\neg\textsc{TryCandidates}(X,\;f,\;W,\;T)$
         \textbf{then break}
  \STATE $\textsc{MiniExplore}(f,\;W,\;T,\;C.m)$
\ENDFOR

\STATE \textbf{return} $\textsc{Coverage}(f)$
\end{algorithmic}
\end{algorithm}

\textsc{TryCandidates} executes each syntactically valid LLM candidate that is absent from $T$. Each candidate runs concolically, updates $W$ and $T$, and returns true iff at least one execution increases line coverage. \textsc{Coverage}$(f)$ is the same line coverage measure used in the evaluation. It is only a progress signal, not a correctness oracle. \textsc{UpdateStale} returns $0$ after coverage growth and $\mathit{stale}+1$ otherwise. In our configuration, $C.\tau=5$, $C.r=3$, $C.m=20$, and $C.\mathit{maxIter}=50$.

LLM candidates are ordinary concrete inputs. They are checked by concolic execution, added to $T$ only when valid and new, and may add path constraints to $W$. The LLM never marks a path feasible or infeasible. Plateau discovery and discovery after the loop require a coverage increase to continue. Discovery after solver failure may collect constraints even from a candidate that does not improve coverage, but the stale counter resets only on coverage growth.

\subsection{Stage 2: Enhanced Concolic Loop}
\label{sec:stage2}

The main loop receives the constraint pool $W$, already populated by seeds, and the set of tried inputs $T$. Each iteration exits if $W$ is empty or full coverage is reached. If the stale counter reaches $\tau$, the engine flushes $W$ and enters \emph{plateau discovery}. The LLM receives annotated source, CFG paths toward uncovered lines, and recent tried inputs. \textsc{TryCandidates} then checks whether any returned candidate improves coverage. A successful batch resets the stale counter. A batch with no coverage gain stops the loop.

Otherwise, the engine pops a path constraint $\phi$ and calls the SMT solver. \textsc{Unknown} or \textsc{Error} triggers \emph{discovery after solver failure}: the LLM sees the source and current coverage gaps, proposes a concrete input, and concolic execution may add new constraints to $W$. The LLM is not asked to solve the failed formula. \textsc{Unsat} paths are skipped. For \textsc{Sat}, the model is merged with $a_0$, executed if new, and used to update the stale counter by observed line coverage.

After the main loop, a \emph{discovery push after the loop} targets remaining uncovered lines for up to $r$ rounds. An LLM round that improves coverage is followed by \textsc{MiniExplore}, which spends at most $C.m$ solver iterations on constraints collected from those candidates. A round with no improvement stops the push. Thus, the LLM call budget has three parts: one seed call, plateau calls and calls after the loop that require coverage improvement, and calls after solver failure bounded by $C.\mathit{maxIter}$ (50). The mini exploration runs after the main loop for up to $C.r$ rounds only when the last LLM candidate batch improves coverage.

\section{Evaluation}
\label{sec:eval}

We evaluate \tool along three research questions:

\begin{itemize}
\item \textbf{RQ1}: How does \tool compare to \textsc{Concolic-Only}, LLM seed execution without symbolic follow up, and \textsc{CrossHair} in terms of line coverage?
\item \textbf{RQ2}: On which categories of functions does LLM integration provide gains, and where is the solver alone sufficient?
\item \textbf{RQ3}: What is the wall clock time and API cost overhead of LLM integration?
\end{itemize}

Across evaluation tables, CL and CO denote \tool and \textsc{Concolic-Only}, respectively.
LO and CH denote \textsc{LLM-Seeds-Only} and \textsc{CrossHair}, respectively.
All experiments ran on an AMD Ryzen 7700 CPU with 32\,GB RAM. Each run uses \code{max\_iterations=50} and a 90 second watchdog for each target subprocess. The SMT backend is \textsc{CVC5} 1.1.2 with a 15 second timeout per constraint. The LLM backend is \code{gpt-4o-mini-2024-07-18} at temperature 0.7. Reported time is total campaign time, including controller overhead, LLM latency, subprocess management, and checkpoint recovery, so some means can exceed 90 seconds. \tool also includes bounded candidate executions triggered by the LLM and up to three discovery rounds after the loop, each followed by at most 20 mini exploration solver iterations only after an LLM round that improves coverage. Results are averaged over $N = 10$ trials. The artifact, including benchmarks and scripts, will be made public upon acceptance.

\begin{table}[!t]
  \caption{Benchmark suite. Each function accepts primitive or structured inputs and returns a string classification. LoC counts the function body only.}
  \label{tab:benchmarks}
  \centering
  \scriptsize
  \setlength{\tabcolsep}{2pt}
  \renewcommand{\arraystretch}{1.05}
  \begin{tabularx}{\textwidth}{@{}
      >{\raggedright\arraybackslash}p{2.30cm}
      c
      >{\raggedright\arraybackslash}p{4.05cm}
      r
      >{\raggedright\arraybackslash}X
    @{}}
    \toprule
    \textbf{Category} & \textbf{ID} & \textbf{Function} & \textbf{LoC} & \textbf{Functionality} \\
    \midrule
    \multirow{3}{*}{Pure Numeric}
      & B01 & triangle\_classification      & 39 & Classify triangle by shape and angle type \\
      & B02 & tax\_bracket\_calculator     & 37 & Determine marginal tax bracket by income and filing status \\
      & B03 & bmi\_risk\_classifier         & 38 & Compute BMI and map to WHO risk category \\
    \midrule
    \multirow{4}{*}{String Constr.}
      & B04 & email\_validation             & 90 & Validate email format, TLD classification, disposable detection \\
      & B05 & url\_routing                  & 104 & Route URL by protocol, host, path segments, and query params \\
      & B06 & semver\_parsing               & 86 & Parse semantic version with prerelease and build metadata \\
      & B07 & log\_level\_routing           & 88 & Detect log format (JSON/syslog/CSV/plain) and route by severity \\
    \midrule
    \multirow{3}{*}{Library Black Box}
      & B08 & json\_config\_validation      & 91 & Parse JSON and validate keys, types, nested settings \\
      & B09 & regex\_data\_extraction       & 80 & Detect phone/email/date/IP/UUID via regex in free text \\
      & B10 & datetime\_classification      & 92 & Parse dates across ISO/US/EU formats, classify by calendar \\
    \midrule
    \multirow{2}{*}{Hash/Encoding}
      & B11 & credit\_card\_validation      & 92 & Validate card number by prefix, length, and Luhn checksum \\
      & B12 & base64\_payload\_classif.     & 74 & Decode base64 and classify content (JSON/XML/CSV/binary) \\
    \midrule
    \multirow{4}{*}{Mixed Synergy}
      & B13 & http\_request\_classif.       & 62 & Classify HTTP request by method, path, and content length \\
      & B14 & discount\_engine              & 69 & Apply coupon codes and quantity-based bulk discounts \\
      & B15 & access\_control\_checker      & 56 & Decide role/resource/trust-level access permissions \\
      & B16 & shipping\_rate\_calculator    & 56 & Classify shipment by weight, zone, and speed into rate tiers \\
    \midrule
    \multirow{2}{*}{Complex Struct.}
      & B17 & nested\_config\_validator     & 78 & Validate multi-level config dict with cross-field dependencies \\
      & B18 & transaction\_ledger\_analysis & 72 & Analyze transaction list with running balance and anomaly detection \\
    \midrule
    \multirow{2}{*}{Deep Path Dep.}
      & B19 & state\_machine\_validator     & 52 & Validate event sequence against order lifecycle state machine \\
      & B20 & multi\_stage\_form\_valid.    & 82 & Validate pipe-delimited form data through 4 sequential stages \\
    \midrule
    \multirow{2}{*}{Solver-Hard}
      & B21 & string\_similarity\_classif.  & 59 & Compare strings by sorting, reversal, prefix, and edit distance \\
      & B22 & pattern\_matching\_disp.      & 74 & Match glob-style patterns with wildcards and character classes \\
    \bottomrule
  \end{tabularx}
  \par\smallskip
  \footnotesize
  22 functions, 37--104 LoC (avg.\ 70). Categories reflect common failure modes in concolic testing~\cite{PyCT, cadar2008klee, stephens2016driller}. 
\end{table}

\subsection{Architecture}
\label{sec:impl}

\tool is implemented as an LLM plugin on top of the \textsc{PyCT}~\cite{PyCT} concolic engine. \textsc{PyCT} still handles symbolic execution, constraint solving, and coverage tracking. The plugin exposes three LLM roles: seed generation, discovery after solver failure, and plateau discovery, with plateau discovery dispatched both in the loop and after the loop.

Prompts include the target source, parameter types, relevant callees, CFG paths toward uncovered code, coverage gaps, and recent tried inputs when available. The model must return a Python list of input dictionaries. Invalid outputs are discarded. Seed lists are generated once and reused by both \tool and \textsc{LLM-Seeds-Only}. Measurement is separated from instrumentation: each target is executed once under concolic proxies to collect constraints and once under primitive inputs to measure line coverage. A parent watchdog preserves the latest completed checkpoint if a target exceeds the execution budget.

\subsection{Benchmarks}

Table~\ref{tab:benchmarks} lists our controlled suite of 22 designed functions. They are grouped by their main challenge. The suite includes numeric logic that is friendly to the solver, string and format constraints, library calls, structured inputs, and path dependent behavior. In addition to the controlled suite, we evaluate 22 functions from six open source Python libraries (Table~\ref{tab:realworld}). These targets exercise similar challenges in production code and were not designed around any tool.

%
%
%
%
%
\begin{table}[!t]
  \caption{Real-world benchmark results.}
  \label{tab:realworld}
  \centering
  \scriptsize
  \setlength{\tabcolsep}{4pt}
	\begin{tabular*}{\textwidth}{@{\extracolsep{\fill}}
      >{\raggedright\arraybackslash}p{2.25cm}
      >{\raggedright\arraybackslash}p{2.90cm}
      r r r r r
    @{}}
    \toprule
    \textbf{Library} & \textbf{Function} & \textbf{LoC}
      & \textbf{CL} & \textbf{CO} & \textbf{LO} & \textbf{CH} \\
    \midrule
    \multirow{6}{*}{werkzeug}
      & options\_hdr & 94 & \textbf{95.2} & 50.0 & 81.9 & 59.6 \\
      & range\_hdr & 50 & \textbf{93.1} & 82.7 & 81.7 & 0.0 \\
      & content\_range & 44 & \textbf{90.0} & 61.3 & 86.1 & 19.4 \\
      & dict\_hdr & 45 & \textbf{96.5} & 74.2 & 94.8 & 83.9 \\
      & list\_hdr & 44 & \textbf{100} & 85.7 & \textbf{100} & \textbf{100} \\
      & cookie & 28 & \textbf{72.1} & 69.0 & 71.7 & 0.0 \\
    \midrule
    \multirow{8}{*}{validators}
      & url & 84 & \textbf{62.2} & 52.4 & 55.1 & 0.0 \\
      & email & 77 & \textbf{56.1} & 37.9 & 48.6 & 34.5 \\
      & hostname & 76 & \textbf{60.5} & 53.5 & 54.2 & 0.0 \\
      & domain & 49 & \textbf{54.2} & 41.8 & 51.6 & 0.0 \\
      & ru\_inn & 50 & \textbf{53.2} & 41.9 & 51.6 & \textbf{53.2} \\
      & fr\_ssn & 40 & \textbf{47.7} & 35.4 & \textbf{47.7} & 0.0 \\
      & ipv4 & 49 & \textbf{48.9} & 41.0 & 48.0 & 42.6 \\
      & country\_code & 43 & 48.8 & \textbf{49.2} & 43.1 & 0.0 \\
    \midrule
    \multirow{2}{*}{phone\-numbers}
      & parse & 130 & \textbf{78.5} & 64.1 & 52.9 & 0.0 \\
      & possible\_num & 25 & \textbf{76.4} & 47.6 & 54.4 & 0.0 \\
    \midrule
    \multirow{4}{*}{urllib.parse}
      & urlsplit & 49 & \textbf{75.9} & 72.9 & 20.7 & 0.0 \\
      & quote & 44 & \textbf{80.3} & 71.1 & 61.3 & 17.4 \\
      & urlencode & 60 & \textbf{79.4} & 52.2 & 60.8 & 14.1 \\
      & parse\_qs & 32 & \textbf{73.8} & 62.1 & 73.1 & 0.0 \\
    \midrule
    simplejson & loads & 65 & \textbf{80.7} & 55.2 & 64.2 & 25.4 \\
    dateutil & parse & 79 & \textbf{64.1} & 53.2 & 57.4 & 0.0 \\
    \midrule
    \multicolumn{2}{@{}l}{\textbf{Average}}
      & & \textbf{72.2} & 57.0 & 61.9 & 20.5 \\
    \bottomrule
  \end{tabular*}
  \par\smallskip
  \footnotesize
  Coverage is mean line coverage (\%) over $N=10$ trials for 22 functions from six libraries. Function names omit repeated package prefixes. CH=0 means budget exhaustion, which occurred on 55\,\% of target runs.
\end{table}

To evaluate at library scale, we use 75 automatically selected entry points from \code{sympy.ntheory} and {PyYAML}'s \code{yaml} package. A target is included if it is defined in the package, accepts at most five parameters, and its parameters can be mapped to primitive values or primitive containers. This mapping uses annotations, defaults, local literal comparisons, or a neutral fallback. Class constructors are included through their \code{\_\_init\_\_} signatures. APIs that require arbitrary user-defined objects are excluded. The suite contains every callable in the two packages that satisfies these criteria. This gives a less curated test of whether the architecture transfers beyond hand picked challenges. Table~\ref{tab:library_summary} reports coverage at the package level for this suite.

\begin{table}[!t]
  \caption{Library-suite results.}
  \label{tab:library_summary}
  \centering
  \scriptsize
  \setlength{\tabcolsep}{4pt}
  \begin{tabular*}{\textwidth}{@{\extracolsep{\fill}} l r r r r r r r r @{}}
    \toprule
    \textbf{Package} & \textbf{\textit{n}}
      & \textbf{CL} & \textbf{CO} & \textbf{LO} & \textbf{CH}
      & $\boldsymbol{\Delta}$
      & $\boldsymbol{p}$ & $\hat{\boldsymbol{A}}_{12}$ \\
    \midrule
    \code{sympy.ntheory} & 37
      & \textbf{79.9} & 55.6 & 78.6 & 36.6
      & $+24.3$ & $8.8\!\times\!10^{-6}$ & 0.71 \\
    {PyYAML} (\code{yaml}) & 38
      & \textbf{77.4} & 67.6 & 75.9 & 35.8
      & $+9.8$ & $2.4\!\times\!10^{-2}$ & 0.53 \\
    \midrule
    \textbf{Overall} & \textbf{75}
      & \textbf{78.7} & 61.7 & 77.3 & 36.2
      & $+17.0$ & $4.1\!\times\!10^{-7}$ & 0.63 \\
    \bottomrule
  \end{tabular*}
  \par\smallskip
  \footnotesize
  Coverage is mean line coverage (\%) by package. $\Delta$ = CL - CO in percentage points. $p$ and $\hat A_{12}$ summarize CL vs.\ CO.
\end{table}

\subsection{Comparison Tools}

We compare \tool with three baselines on the same target set and external time budget. Its bounded candidate executions triggered by the LLM and mini exploration after the loop are part of the proposed policy and are analyzed as overhead in RQ3:

\begin{enumerate}
\item \textsc{Concolic-Only}~\cite{PyCT}, the base \textsc{PyCT} engine with no extensions. This isolates the SMT solver's contribution.
\item \textsc{LLM-Seeds-Only}, which executes the same generated seed inputs concretely, with no concolic replay, constraint collection, or solver follow up. It isolates the shared seed generator and is not a full iterative LLM test generator.
\item \textsc{CrossHair}~\cite{crosshair}, an independent Python symbolic execution tool using \textsc{Z3}. 
\end{enumerate}

\subsection{Statistical Analysis}

For each target and runner, we average line coverage over $N=10$ trials. Pairwise significance tests use one paired observation per target: the mean at the target level for \tool and the corresponding comparator mean. We use the one-sided Wilcoxon signed-rank test~\cite{wilcoxon1945}. Table~\ref{tab:wtl} reports wins, ties, and losses across target runs only descriptively. Those counts are not independent samples for significance testing.

We also report the Vargha-Delaney $\hat A_{12}$ effect size~\cite{vargha2000critique}, which summarizes how often \tool obtains higher coverage than the comparator under the table's paired aggregation. We use it because coverage is bounded and non-normal.

\subsection{RQ1: Coverage Improvement}
\label{sec:rq1}

Table~\ref{tab:main_coverage_dist} summarizes line coverage across the three evaluation suites. It includes distributional statistics and pairwise significance tests. Table~\ref{tab:wtl} breaks down the comparison at the level of target runs. Table~\ref{tab:coverage} reports line coverage for each function in the standard suite.

\begin{table}[!t]
  \caption{Coverage across evaluation suites.}
  \label{tab:main_coverage_dist}
  \centering
  \scriptsize
  \setlength{\tabcolsep}{4pt}
  \begin{tabular*}{\textwidth}{@{\extracolsep{\fill}} l r r r r r r @{}}
    \toprule
    \textbf{Suite} & \textbf{n}
      & \textbf{CL}
      & \textbf{CO}
      & \textbf{LO}
      & $\boldsymbol{p}$
      & $\hat{\boldsymbol{A}}_{12}$ \\
    \midrule
    Standard  & 22  & \textbf{93.2} & 84.6 & 81.7 & $6.5\!\times\!10^{-2}$ & 0.60 \\
    Realworld & 22  & \textbf{72.2} & 57.0 & 61.9 & $4.9\!\times\!10^{-5}$ & 0.75 \\
    Library   & 75  & \textbf{78.7} & 61.7 & 77.3 & $4.1\!\times\!10^{-7}$ & 0.63 \\
    \midrule
    \textbf{Pooled} & 119 & \textbf{78.5} & 64.0 & 74.1 & $<\!10^{-10}$ & 0.66 \\
    \bottomrule
  \end{tabular*}
  \par\smallskip
  \footnotesize
  Values are mean line coverage (\%) over $N=10$ trials. $p$ and $\hat A_{12}$ summarize CL vs.\ CO. \textsc{CrossHair} is omitted for space.
\end{table}

\begin{table}[!t]
  \caption{Pairwise outcomes by target run.}
  \label{tab:wtl}
  \centering
  \scriptsize
  \setlength{\tabcolsep}{4pt}
  \begin{tabular*}{\textwidth}{@{\extracolsep{\fill}} l r c c c c c c @{}}
    \toprule
    \multirow{2}{*}{\textbf{Suite}} & \multirow{2}{*}{\textbf{n}}
      & \multicolumn{3}{c}{\textbf{CL vs.\ CO}}
      & \multicolumn{3}{c}{\textbf{CL vs.\ LO}} \\
    \cmidrule(lr){3-5} \cmidrule(lr){6-8}
      & & \textbf{W} & \textbf{T} & \textbf{L}
      & \textbf{W} & \textbf{T} & \textbf{L} \\
    \midrule
    Standard  &  220 & 61\,\% & 29\,\% & 10\,\% & 60\,\% & 40\,\% &  0\,\% \\
    Realworld &  220 & 96\,\% &  1\,\% &  3\,\% & 69\,\% & 31\,\% &  0\,\% \\
    Library   &  750 & 52\,\% & 44\,\% &  4\,\% & 18\,\% & 82\,\% &  0\,\% \\
    \midrule
    \textbf{Pooled} & 1190 & \textbf{62\,\%} & \textbf{33\,\%} & \textbf{5\,\%}
      & \textbf{35\,\%} & \textbf{65\,\%} & \textbf{0\,\%} \\
    \bottomrule
  \end{tabular*}
  \par\smallskip
  \footnotesize
  W/T/L compares each target run under the stated budget. Counts are descriptive and are not independent samples for signed-rank tests.
\end{table}

\begin{table}[!t]
  \caption{Standard-suite per-benchmark results.}
  \label{tab:coverage}
  \centering
  \scriptsize
  \setlength{\tabcolsep}{4pt}
  \begin{tabular*}{\textwidth}{@{\extracolsep{\fill}} >{\centering\arraybackslash}p{1.4cm} r r r r r r r r r @{}}
    \toprule
    & \multicolumn{4}{c}{\textbf{Coverage (\%)}}
      & $\boldsymbol{\Delta}$
      & \multicolumn{4}{c}{\textbf{Time (s)}} \\
    \cmidrule(lr){2-5} \cmidrule(lr){7-10}
    \textbf{ID}
      & \textbf{CL}
      & \textbf{CO}
      & \textbf{LO}
      & \textbf{CH}
      &
      & \textbf{CL}
      & \textbf{CO}
      & \textbf{LO}
      & \textbf{CH} \\
    \midrule
    \multicolumn{10}{@{}l}{\textit{Pure Numeric}} \\
    B01 & \textbf{100} & \textbf{100} & \textbf{100} & \textbf{100} & 0 & 5.0 & 0.5 & 4.7 & 31.6 \\
    B02 & \textbf{100} & \textbf{100} & \textbf{100} & \textbf{100} & 0 & 6.7 & 0.4 & 6.4 & 3.0 \\
    B03 & 86.7 & \textbf{100} & 78.9 & \textbf{100} & $-13.3$ & 10.7 & 0.5 & 3.6 & 45.2 \\
    \midrule
    \multicolumn{10}{@{}l}{\textit{String Constraints}} \\
    B04 & \textbf{92.3} & 88.5 & 82.7 & --- & $+3.8$ & 112.8 & 101.9 & 2.8 & --- \\
    B05 & \textbf{100} & 96.0 & 90.4 & --- & $+4.0$ & 3.7 & 92.1 & 3.1 & --- \\
    B06 & \textbf{98.1} & 87.5 & 80.0 & 46.9 & $+10.6$ & 15.1 & 0.9 & 2.3 & 11.1 \\
    B07 & \textbf{89.3} & 86.7 & 82.7 & --- & $+2.7$ & 95.5 & 13.9 & 2.7 & --- \\
    \midrule
    \multicolumn{10}{@{}l}{\textit{Library Black Box}} \\
    B08 & \textbf{86.7} & 79.2 & 52.1 & 10.8 & $+7.5$ & 14.6 & 0.6 & 4.0 & 95.4 \\
    B09 & \textbf{100} & 31.6 & \textbf{100} & --- & $+68.4$ & 3.4 & 0.3 & 3.1 & --- \\
    B10 & \textbf{97.0} & 90.0 & 95.0 & 55.0 & $+7.0$ & 8.6 & 0.4 & 2.3 & 13.8 \\
    \midrule
    \multicolumn{10}{@{}l}{\textit{Hash/Encoding}} \\
    B11 & \textbf{100} & \textbf{100} & 91.1 & --- & 0 & 12.9 & 2.5 & 3.5 & --- \\
    B12 & \textbf{84.5} & 54.5 & 75.5 & 31.8 & $+30.0$ & 15.3 & 0.4 & 7.0 & 8.0 \\
    \midrule
    \multicolumn{10}{@{}l}{\textit{Mixed Type Synergy}} \\
    B13 & \textbf{100} & \textbf{100} & \textbf{100} & 33.3 & 0 & 4.2 & 0.6 & 3.9 & 6.7 \\
    B14 & \textbf{100} & \textbf{100} & \textbf{100} & \textbf{100} & 0 & 4.6 & 0.6 & 4.3 & 17.6 \\
    B15 & 77.0 & \textbf{100} & 36.0 & 30.0 & $-23.0$ & 16.5 & 0.6 & 6.3 & 4.7 \\
    B16 & 87.5 & \textbf{100} & 56.7 & 58.3 & $-12.5$ & 10.8 & 0.6 & 5.0 & 4.2 \\
    \midrule
    \multicolumn{10}{@{}l}{\textit{Complex Structures}} \\
    B17 & \textbf{72.9} & 61.9 & 39.0 & --- & $+11.0$ & 10.4 & 0.3 & 4.8 & --- \\
    B18 & \textbf{91.5} & 46.2 & 70.8 & 46.2 & $+45.4$ & 10.4 & 0.3 & 4.8 & 12.3 \\
    \midrule
    \multicolumn{10}{@{}l}{\textit{Deep Path Dependency}} \\
    B19 & \textbf{91.7} & 83.3 & 84.2 & 58.3 & $+8.3$ & 9.0 & 0.8 & 1.8 & 5.7 \\
    B20 & \textbf{100} & 89.5 & 95.3 & 78.9 & $+10.5$ & 9.3 & 2.8 & 4.2 & 18.8 \\
    \midrule
    \multicolumn{10}{@{}l}{\textit{Solver-Hard}} \\
    B21 & \textbf{98.8} & 81.2 & 91.9 & --- & $+17.5$ & 23.3 & 103.5 & 2.5 & --- \\
    B22 & 96.9 & 84.6 & 95.4 & \textbf{100} & $+12.3$ & 31.1 & 17.6 & 3.1 & 34.5 \\
    \midrule
    \textbf{Avg}
      & \textbf{93.2}
      & 84.6
      & 81.7
      & 44.0
      & $+8.6$
      & 19.7
      & 15.5
      & 3.9
      & 45.8 \\
    \bottomrule
  \end{tabular*}
  \par\smallskip
  \footnotesize
  Coverage and time are means over $N=10$ trials. Bold marks best coverage. ``---'' means timeout. IDs map to Table~\ref{tab:benchmarks}.
\end{table}

Across the three suites (Table~\ref{tab:main_coverage_dist}), \tool improves mean coverage over \textsc{Concolic-Only} by $+8.6$, $+15.1$, and $+17.0$\,pp. The evidence is strongest on real-world and library targets. The gain on the controlled suite is positive but below the 0.05 threshold, partly because many functions already let the solver reach saturation.

Table~\ref{tab:wtl} shows that wins dominate losses against \textsc{Concolic-Only}. The visible regressions, B03, B15, and B16, come from seed replay changing exploration order under a fixed iteration budget. Against \textsc{LLM-Seeds-Only}, the pattern matches the intended division of labor: generated inputs open useful states, and concolic replay explores variants from those states.

The library suite shows the same mechanism at scale. \textsc{Concolic-Only} is bimodal: some targets saturate, while others stall behind semantic or library barriers. \tool moves many stalled cases upward. The {PyYAML} result is statistically detectable but practically small by $\hat A_{12}$, so the stronger evidence at the package level comes from \code{sympy.ntheory} and from the library suite overall. Gains on real code follow the same explanation on regex, tokenizer, parser, and deep library gates.

\begin{table}[!t]
  \caption{Post hoc dominance on individually reported targets.}
  \label{tab:target_dominance_posthoc}
  \centering
  \scriptsize
  \setlength{\tabcolsep}{4pt}
  \begin{tabular*}{\textwidth}{@{\extracolsep{\fill}} l c c c c @{}}
    \toprule
    \textbf{Suite} & \textbf{n} & \textbf{vs.\ CO} & \textbf{vs.\ LO} & \textbf{vs.\ max(CO, LO)} \\
    \midrule
    Standard & 22 & 14 / 5 / 3 & 17 / 5 / 0 & 13 / 6 / 3 \\
    Real-world & 22 & 21 / 0 / 1 & 20 / 2 / 0 & 19 / 2 / 1 \\
    Total & 44 & 35 / 5 / 4 & 37 / 7 / 0 & 32 / 8 / 4 \\
    \bottomrule
  \end{tabular*}
  \par\smallskip
  \footnotesize
  Entries show CL $>$ / $=$ / $<$ counts computed from the mean coverage values already reported for each target in Tables~\ref{tab:coverage} and~\ref{tab:realworld}. $\max(\mathrm{CO},\mathrm{LO})$ denotes the better single baseline on that target. The library suite is omitted because the manuscript reports means at the package level there.
\end{table}

To distinguish integration gains from whichever single baseline happens to be stronger on a target, we performed a post hoc dominance check on the 44 individually reported functions using the mean coverage values already reported for each target. \tool exceeds \textsc{Concolic-Only} on 35/44 targets and exceeds \textsc{LLM-Seeds-Only} on 37/44. More importantly, it exceeds the better of the two single baselines on 32/44 targets, ties on 8/44, and trails on only 4/44. The losses are confined to the already discussed cases where the solver already suffices or to cases where seeds reordered exploration (B03, B15, B16), plus one near tie on the real-world validator (\texttt{country\_code}). This is still a coarse integration comparison rather than an ablation of individual components, but it strengthens the interpretation that the main benefit usually comes from combining concrete inputs that match the domain with concolic replay, not from either component alone.

\textbf{RQ1 verdict.} \tool improves mean line coverage over \textsc{Concolic-Only} in all three evaluation suites, with the strongest statistical and practical evidence on real-world and library targets. The controlled suite shows a positive but weaker mean gain because many functions already let the solver reach saturation. The comparison with \textsc{LLM-Seeds-Only}, together with the post hoc dominance check at the target level, shows that generated concrete examples alone do not explain the full result. Their value is greatest when concolic replay can collect constraints and explore variants from the states they reach.

\subsection{RQ2: When Does the LLM Help?}

\begin{table}[!t]
  \caption{Coverage by controlled suite category.}
  \label{tab:category}
  \centering
  \scriptsize
  \setlength{\tabcolsep}{4pt}
  \begin{tabular*}{\textwidth}{@{\extracolsep{\fill}} >{\raggedright\arraybackslash}p{3.25cm} r r r r r @{}}
    \toprule
    \textbf{Category}
      & \textbf{CL}
      & \textbf{CO}
      & \textbf{LO}
      & $\boldsymbol{p}$
      & $\hat{\boldsymbol{A}}_{12}$ \\
    \midrule
    Pure Numeric    &  95.6 & 100.0 &  93.0 & --- & 0.33 \\
    String Constr.  &  94.9 &  89.7 &  83.9 & 0.10 & 0.88 \\
    Library B.B.    &  94.6 &  66.9 &  82.4 & 0.18 & 0.89 \\
    Hash/Encoding   &  92.3 &  77.3 &  83.3 & --- & 0.62 \\
    Mixed Synergy   &  91.1 & 100.0 &  73.2 & --- & 0.25 \\
    Complex Struct. &  82.2 &  54.0 &  54.9 & --- & 1.00 \\
    Deep Path Dep.  &  95.8 &  86.4 &  89.7 & --- & 1.00 \\
    Solver-Hard     &  97.8 &  82.9 &  93.6 & --- & 1.00 \\
    \midrule
    \textbf{Overall} & \textbf{93.2} & 84.6 & 81.7 & $6.5\!\times\!10^{-2}$ & 0.60 \\
    \bottomrule
  \end{tabular*}
  \par\smallskip
  \footnotesize
  Values are category mean coverage (\%). Library B.B. = Library Black Box. $p$ and $\hat A_{12}$ summarize CL vs.\ CO. ``---'' means too few targets.
\end{table}

Table~\ref{tab:category} reports average coverage by failure mode category.
The category breakdown shows two regimes. When the solver is already sufficient, as in Pure Numeric and Mixed Synergy targets, LLM seed replay can hurt. It spends early budget on concrete examples before the solver reaches its usual saturation path. The reproducible regressions appear in three controlled targets. There, plausible but incomplete LLM seeds stop validation at the first branch and delay later path discovery. When the solver is limited, as in library black boxes, complex structures, hash/encoding checks, and constraints the solver cannot handle well, the pattern reverses. Concrete LLM seeds cross barriers that pure constraint reasoning cannot model.

The three runners provide a coarse integration comparison. \textsc{Concolic-Only} isolates search by the solver. \textsc{LLM-Seeds-Only} measures how much coverage the shared generated seeds achieve without symbolic exploration. \tool measures the complete proposed policy: seed generation, concolic replay, discovery after solver failure, plateau discovery, and discovery after the loop. They show that the full integration is strongest when concrete inputs generated by the LLM open states from which concolic execution can collect constraints and explore variants.

\begin{table}[!t]
  \caption{Follow-up additive ablation of selected LLM roles.}
  \label{tab:ablation_followup}
  \centering
  \scriptsize
  \setlength{\tabcolsep}{4pt}
  \begin{tabular*}{\textwidth}{@{\extracolsep{\fill}} l r r r @{}}
    \toprule
    \textbf{Comparison} & \textbf{Standard} & \textbf{Library} & \textbf{Real-world} \\
    \midrule
    + Seeds vs.\ CO & $+6.6$ & $+14.9$ & $+9.3$ \\
    + Plateau vs.\ CO & $+10.6$ & $+14.5$ & $+9.2$ \\
    Full system vs.\ CO & $+9.5$ & $+16.6$ & $+13.0$ \\
    Full system vs.\ LO & $+10.5$ & $+8.3$ & $+13.6$ \\
    \bottomrule
  \end{tabular*}
  \par\smallskip
  \footnotesize
  Values are line coverage deltas in percentage points from a separate $N=10$ follow-up rerun, not replacements for the main coverage numbers. The first three rows compare with \textsc{Concolic-Only} in that rerun. The last row compares the full system with LO.
\end{table}

Table~\ref{tab:ablation_followup} adds a component-level view. Seed generation and plateau discovery both give large standalone gains over \textsc{Concolic-Only}, while the full system remains clearly above \textsc{LLM-Seeds-Only}. This supports the intended division of labor: the LLM is useful for reaching concrete regions that the solver misses, but concolic replay is still needed to turn those regions into systematic exploration. The ablation is additive rather than leave-one-out, and the plateau row combines the in-loop trigger and the push after the loop, so it should be read as evidence about dominant roles rather than exact marginal contributions.

Table~\ref{tab:rw_category} reports average coverage by real-world challenge category. The pattern mirrors the standard suite but with sharper contrasts.
Descriptively, \tool exceeds both CO and LO in 10 of the 12 category means across the controlled and real-world breakdowns. The two exceptions are Pure Numeric and Mixed Synergy, which are precisely the regime where the solver already suffices.

\begin{table}[!t]
  \caption{Coverage by real-world challenge category.}
  \label{tab:rw_category}
  \centering
  \scriptsize
  \setlength{\tabcolsep}{4pt}
  \begin{tabular*}{\textwidth}{@{\extracolsep{\fill}} >{\raggedright\arraybackslash}p{3.35cm} r r r r r r @{}}
    \toprule
    \textbf{Category} (\textit{n})
      & \textbf{CL}
      & \textbf{CO}
      & \textbf{LO}
      & $\boldsymbol{\Delta}$
      & $\boldsymbol{p}$
      & $\hat{\boldsymbol{A}}_{12}$ \\
    \midrule
    Format/Regex (12)  &  70.0 &  54.3 &  65.8 & $+15.7$ & --- & 0.90 \\
    Opaque Lib. Disp. (4) &  67.2 &  53.5 &  54.8 & $+13.7$ & --- & 1.00 \\
    Mostly Solver-Tr. (4) &  76.6 &  63.4 &  52.4 & $+13.2$ & --- & 0.65 \\
    Non-Primitive (2)  &  86.2 &  67.5 &  71.3 & $+18.8$ & --- & 0.60 \\
    \midrule
    \textbf{Overall (22)} & \textbf{72.2} & 57.0 & 61.9 & $+15.1$ & $\!\!\!4.9\!\times\!10^{-5}$ & 0.75 \\
    \bottomrule
  \end{tabular*}
  \par\smallskip
  \footnotesize
  Values are category mean coverage (\%). $\Delta$ = CL - CO in percentage points. $p$ and $\hat A_{12}$ summarize CL vs.\ CO. ``---'' means too few targets.
\end{table}

Format/regex and opaque dispatch targets show the clearest division of labor. The LLM supplies strings shaped to the domain, JSON structures, phone number formats, and parser inputs that pass gates outside the solver's model. Concolic replay then explores the Python level branches exposed by those examples. The common pattern is CL\,$>$\,LO\,$>$\,CO: seed generation opens the region, and solver negation inside that region adds coverage.

The limits appear on mostly solver tractable and non primitive targets. If \textsc{Concolic-Only} would saturate quickly, the early seed phase can reorder exploration and consume part of the fixed budget, as in B03, B15, and B16. If an API requires dictionaries or structured objects that \textsc{PyCT} cannot proxy symbolically, coverage depends heavily on concrete input quality. Table~\ref{tab:co_bucket} summarizes the trend: gains are largest below $60$\,\% CO coverage and mostly disappear above $85$\,\%, unless a late semantic gate remains.

This pattern suggests a practical rule of thumb. \tool is most useful when the current frontier is blocked by semantic checks that are easy for developers to recognize but hard for the SMT solver to model, such as parsers, regular expressions, checksums, and structured input formats. In these cases, an LLM candidate can cross the gate, and concolic replay can recover systematic exploration on the newly reached Python code. The method is less useful when the remaining branches are already solver tractable or when the API requires objects outside the current concolic input model. Then LLM seeds mainly change exploration order or depend on concrete input quality rather than opening a new symbolic frontier.

\begin{table}[!t]
  \caption{Coverage gains grouped by \textsc{Concolic-Only} baseline coverage.}
  \label{tab:co_bucket}
  \centering
  \scriptsize
  \setlength{\tabcolsep}{4pt}
  \begin{tabular*}{\textwidth}{@{\extracolsep{\fill}} l r r r r r @{}}
    \toprule
    \textbf{Suite}
      & \textbf{CO $\in [0,30)$}
      & \textbf{$[30,60)$}
      & \textbf{$[60,85)$}
      & \textbf{$[85,100]$}
      & \textbf{Overall} \\
    \midrule
    Standard  & ---     & $+47.9$ & $+11.3$ & $-0.7$  & $+8.6$  \\
    Realworld & ---     & $+16.6$ & $+12.9$ & $+14.3$ & $+15.1$ \\
    Library   & $+29.7$ & $+40.6$ & $+18.2$ & $+0.6$  & $+17.0$ \\
    \bottomrule
  \end{tabular*}
  \par\smallskip
  \footnotesize
  Cells show the mean CL - CO coverage gap (pp) for target trials in each CO bucket. ``---'' means none.
\end{table}

\textbf{RQ2 verdict.} The LLM helps when coverage is blocked by formats, parsers, library dispatch, or structured inputs, not ordinary arithmetic constraints. Its main role is to reach useful concrete regions. The solver still performs systematic exploration inside those regions. The follow-up ablation suggests that seeds and plateau discovery carry most of the observed gain, although finer leave-one-out ablations are still needed to measure interactions among the roles.

\subsection{RQ3: Cost and Runtime Efficiency}
\label{sec:rq3}

\begin{table}[!t]
  \caption{LLM cost and runtime per target trial.}
  \label{tab:cost}
  \centering
  \scriptsize
  \setlength{\tabcolsep}{4pt}
  \begin{tabular*}{\textwidth}{@{\extracolsep{\fill}} l r r r r r r @{}}
    \toprule
    \multirow{2}{*}{\textbf{Suite}}
      & \multicolumn{3}{c}{\textbf{LLM cost}}
      & \multicolumn{3}{c}{\textbf{Time (s)}} \\
    \cmidrule(lr){2-4} \cmidrule(lr){5-7}
      & \textbf{Tokens} & \textbf{USD} & \textbf{Tok/pp}
      & \textbf{CL} & \textbf{CO} & \textbf{LO} \\
    \midrule
    Standard  &  1{,}548 & 0.0004 & 180 & 19.7 & 15.5 & 3.9 \\
    Realworld & 11{,}310 & 0.0021 & 750 & 35.0 &  1.9 & 5.9 \\
    Library   &  7{,}868 & 0.0014 & 460 & 24.2 &  1.8 & 4.6 \\
    \bottomrule
  \end{tabular*}
  \par\smallskip
  \footnotesize
  Tokens, USD, and time are averages per target trial. Tok/pp = tokens per coverage point over CO. Time is total wall clock time, including LLM calls and controller overhead. Dollar costs use \code{gpt-4o-mini-2024-07-18} rates.
\end{table}

\textbf{Token cost.}
LLM use scales with target complexity. Controlled targets average 1{,}548 tokens per trial, while real-world targets average 11{,}310 because prompts are longer and plateau calls and calls after the loop are more frequent. The full evaluation over 1{,}190 target runs costs \$1.63 in API charges at the \code{gpt-4o-mini-2024-07-18} rates used during the experiment.

\textbf{Wall clock time.}
\tool is modestly slower on the controlled suite, where \textsc{Concolic-Only} also spends time in solver calls (19.7\,s vs.\ 15.5\,s). On real-world and library suites, \tool is roughly 13--18$\times$ slower because total time includes LLM calls, controller overhead, subprocess management, and checkpoint recovery. The main overheads are seed generation, subprocess forks, plateau calls in the loop, and discovery calls after the loop. \tool remains faster than \textsc{CrossHair} on many real-world targets where \textsc{CrossHair} often times out, but \textsc{CrossHair}'s timeout behavior makes mean runtime comparisons difficult.

\textbf{Cost scaling.}
The LLM call budget has three parts: one seed call, plateau calls and calls after the loop that stop when coverage no longer improves, and calls after solver failure bounded by the main concolic iteration budget. Plateau use and use after the loop stop after a round with no improvement. Use after solver failure scales with \textsc{Unknown}/\textsc{Error} results within \code{max\_iterations}. Thus, incremental cost grows with the number of hard functions, while targets that the solver already saturates usually incur only seed generation cost.

\textbf{RQ3 verdict.} Under the model and prices used in the experiment, dollar cost is small, but latency is material. This makes bounded triggering essential. \tool should be used as a targeted escape hatch for solver stalls and semantic barriers, not as a replacement for ordinary concolic search.

\section{Limitations and Concluding Remarks}
\label{sec:limitations}

\tool shows that an LLM need not replace symbolic execution to help Python concolic testing. Its useful role is narrower: propose concrete inputs that cross semantic barriers, let execution validate them, and let the concolic engine resume systematic exploration from newly reached states. Across the three benchmark groups, the complete policy improves mean line coverage over \textsc{Concolic-Only}, with the strongest gains on regexes, parsers, checksums, structured inputs, and opaque library boundaries. The weakest gains appear when the solver already saturates the target, where seed replay can reorder exploration under a fixed budget.

The results should be read as evidence about input generation, not fault detection. Line coverage is a standard proxy in concolic-testing evaluations~\cite{cadar2008klee, godefroid2008sage, stephens2016driller}, but it does not measure assertion quality, mutation score, real bug detection, or branches that share a line. Future work should test whether the added coverage improves mutation scores, seeded bug detection, real fault detection, and robustness or fairness analyses for learned components~\cite{huang2025pyfair,hong2025transformerrobustness}. \textsc{LLM-Seeds-Only} isolates the shared seed generator, but it is not a substitute for iterative LLM test-suite generators.

The evaluation compares complete policies under fixed budgets. The follow-up ablation identifies dominant roles, but it does not provide leave-one-out marginal effects from the full system. The plateau also combines an in-loop trigger and a push after the loop, and discovery after the loop may add bounded mini exploration after the main loop. \textsc{CrossHair} uses the same external timeout, although treating timeouts as zero can understate its capability on slow targets. The watchdog checkpoint protocol affects measurement preservation only: it returns the latest completed exploration state if a child process is killed.

Finally, LLM output is stochastic, and the benchmark set is broad but not universal. We fix the model snapshot, temperature, budgets, and hardware, and report $N\!=\!10$ means, but results may shift with other models, prompts, rate limits, or hardware. APIs with deep types and callable or object values can leave \tool dominated by concrete input quality, while Python code over primitive values benefits most from concolic follow up. Future work should improve input modeling, run cleaner leave-one-out ablations, test sensitivity to model and provider, and evaluate fault-oriented metrics beyond line coverage.


\begin{thebibliography}{10}
\providecommand{\url}[1]{\texttt{#1}}
\providecommand{\urlprefix}{URL }
\providecommand{\doi}[1]{https://doi.org/#1}

\bibitem{mayhem2018}
Avgerinos, T., Brumley, D., Davis, J., Goulden, R., Nighswander, T., Rebert,
  A., Williamson, N.: The {Mayhem} cyber reasoning system. IEEE Security \&
  Privacy  \textbf{16}(2),  52--60 (2018). \doi{10.1109/MSP.2018.1870873}

\bibitem{PyExZ3}
Ball, T., Daniel, J.: Deconstructing dynamic symbolic execution. In: Dependable
  Software Systems Engineering, NATO Science for Peace and Security Series D,
  vol.~40, pp. 26--41. IOS Press (2015). \doi{10.3233/978-1-61499-495-4-26}

\bibitem{cadar2008klee}
Cadar, C., Dunbar, D., Engler, D.: {KLEE}: Unassisted and automatic generation
  of high-coverage tests for complex systems programs. In: OSDI. USENIX
  Association (2008)

\bibitem{exe2006}
Cadar, C., Ganesh, V., Pawlowski, P.M., Dill, D.L., Engler, D.R.: {EXE}:
  Automatically generating inputs of death. In: CCS. pp. 322--335 (2006).
  \doi{10.1145/1180405.1180445}

\bibitem{cadar2013symbolic}
Cadar, C., Sen, K.: Symbolic execution for software testing: three decades
  later. Communications of the ACM  \textbf{56}(2),  82--90 (2013).
  \doi{10.1145/2408776.2408795}

\bibitem{PyCT}
Chen, Y.F., Tsai, W.L., Wu, W.C., Yen, D.D., Yu, F.: {PyCT}: A {Python}
  concolic tester. In: APLAS. pp. 38--46 (2021).
  \doi{10.1007/978-3-030-89051-3_3}

\bibitem{z3_2008}
De~Moura, L., Bj{\o}rner, N.: {Z3}: An efficient {SMT} solver. In: TACAS. pp.
  337--340. Springer (2008)

\bibitem{deng2023largelanguagemodelszeroshot}
Deng, Y., Xia, C.S., Peng, H., Yang, C., Zhang, L.: Large language models are
  zero-shot fuzzers: Fuzzing deep-learning libraries via large language models.
  arXiv preprint arXiv:2212.14834 (2023). \doi{10.48550/arXiv.2212.14834},
  accepted at ISSTA 2023

\bibitem{sbft2024}
Erni, N., Al-Ameen, M., Birchler, C., Derakhshanfar, P., Lukasczyk, S.,
  Panichella, S.: {SBFT} tool competition 2024 -- {Python} test case generation
  track. In: SBFT. pp. 37--40 (2024). \doi{10.1145/3643659.3643930}

\bibitem{aflpp2020}
Fioraldi, A., Maier, D., Ei{\ss}feldt, H., Heuse, M.: {AFL++}: Combining
  incremental steps of fuzzing research. In: WOOT. USENIX Association (2020)

\bibitem{evosuite2011}
Fraser, G., Arcuri, A.: {EvoSuite}: Automatic test suite generation for
  object-oriented software. In: ESEC/FSE. pp. 416--419 (2011).
  \doi{10.1145/2025113.2025179}

\bibitem{godefroid2005dart}
Godefroid, P., Klarlund, N., Sen, K.: {DART}: Directed automated random
  testing. In: PLDI. pp. 213--223 (2005)

\bibitem{godefroid2008sage}
Godefroid, P., Levin, M.Y., Molnar, D.A.: Automated whitebox fuzz testing. In:
  NDSS. The Internet Society (2008)

\bibitem{hong2025registerautomata}
Hong, C.D., Jiang, H., Lin, A.W., Markgraf, O., Parsert, J., Tan, T.:
  Extracting robust register automata from neural networks over data sequences
  (2025). \doi{10.48550/arXiv.2511.19100}

\bibitem{hong2025transformerrobustness}
Hong, C.D., Yang, C.C., Wang, Y., Yu, F.: Influence-guided concolic testing of
  transformer robustness (2025). \doi{10.48550/arXiv.2509.23806}, accepted at
  the International Conference on Software Quality, Reliability, and Security

\bibitem{huang2025pyfair}
Huang, M.I., Hong, C.D., Yu, F.: Concolic testing on individual fairness of
  neural network models (2025). \doi{10.48550/arXiv.2509.06864}

\bibitem{lemieux2023codamosa}
Lemieux, C., Inala, J.P., Lahiri, S., Sen, S.: {CodaMOSA}: Escaping coverage
  plateaus in test generation with pre-trained large language models. In:
  Proceedings of the 45th IEEE/ACM International Conference on Software
  Engineering (2023)

\bibitem{li2025autobug}
Li, Y., Meng, R., Duck, G.J.: Large language model powered symbolic execution.
  Proc. ACM Program. Lang.  \textbf{9}(OOPSLA2),  3148--3176 (2025).
  \doi{10.1145/3763163}

\bibitem{lin2026rnnrobustness}
Lin, L.J., Hong, C.D.: Robustness verification of recurrent neural networks
  with abstraction refinement (2026). \doi{10.48550/arXiv.2606.12490}

\bibitem{pynguin2022}
Lukasczyk, S., Fraser, G.: {Pynguin}: Automated unit test generation for
  {Python}. In: ICSE Companion. pp. 168--172 (2022).
  \doi{10.1145/3510454.3516829}

\bibitem{luo2026concollmic}
Luo, Z., Zhao, H., Wolff, D., Cadar, C., Roychoudhury, A.: Agentic concolic
  execution. In: IEEE S\&P. pp. 1--19. IEEE Computer Society (2026)

\bibitem{hypothesis2019}
MacIver, D.R., Hatfield-Dodds, Z., Contributors, M.O.: {Hypothesis}: A new
  approach to property-based testing. Journal of Open Source Software
  \textbf{4}(43), ~1891 (2019). \doi{10.21105/joss.01891}

\bibitem{coverup2025}
Pizzorno, J.A., Berger, E.D.: {CoverUp}: Effective high coverage test
  generation for {Python} (2025). \doi{10.1145/3729398}, to appear at FSE 2025

\bibitem{symcc2020}
Poeplau, S., Francillon, A.: Symbolic execution with {SymCC}: Don't interpret,
  compile! In: USENIX Security. pp. 181--198. USENIX Association (2020)

\bibitem{crosshair}
Schanely, P.: {CrossHair}: An analysis tool for {Python}.
  \url{https://github.com/pschanely/CrossHair} (2017)

\bibitem{sen2007concolic}
Sen, K.: Concolic testing. In: Proceedings of the 22nd IEEE/ACM International
  Conference on Automated Software Engineering. pp. 571--572 (2007).
  \doi{10.1145/1321631.1321746}

\bibitem{sen2005cute}
Sen, K., Marinov, D., Agha, G.: {CUTE}: A concolic unit testing engine for {C}.
  In: ESEC/FSE. pp. 263--272 (2005). \doi{10.1145/1081706.1081750}

\bibitem{stephens2016driller}
Stephens, N., Grosen, J., Salls, C., Dutcher, A., Wang, R., Corbetta, J.,
  Shoshitaishvili, Y., Kr{\"u}gel, C., Vigna, G.: {Driller}: Augmenting fuzzing
  through selective symbolic execution. In: NDSS (2016)

\bibitem{tu2026cottontail}
Tu, H., Lee, S., Li, Y., Chen, P., Jiang, L., B{\"o}hme, M.: {COTTONTAIL}:
  Large language model-driven concolic execution for highly structured test
  input generation. arXiv preprint arXiv:2504.17542 (2025).
  \doi{10.48550/arXiv.2504.17542}, to appear at the 2026 IEEE Symposium on
  Security and Privacy

\bibitem{vargha2000critique}
Vargha, A., Delaney, H.D.: A critique and improvement of the {CL} common
  language effect size statistics of {McGraw} and {Wong}. Journal of
  Educational and Behavioral Statistics  \textbf{25}(2),  101--132 (2000).
  \doi{10.3102/10769986025002101}

\bibitem{wang2024llmsym}
Wang, W., Liu, K., Chen, A.R., Li, G., Jin, Z., Huang, G., Ma, L.: {Python}
  symbolic execution with {LLM}-powered code generation. arXiv preprint
  arXiv:2409.09271 (2024). \doi{10.48550/arXiv.2409.09271}

\bibitem{wilcoxon1945}
Wilcoxon, F.: Individual comparisons by ranking methods. Biometrics Bulletin
  \textbf{1}(6),  80--83 (1945). \doi{10.2307/3001968}

\bibitem{yun2018qsym}
Yun, I., Lee, S., Xu, M., Jang, Y., Kim, T.: {QSYM}: A practical concolic
  execution engine tailored for hybrid fuzzing. In: USENIX Security. pp.
  745--761. USENIX Association (2018)

\end{thebibliography}
\end{document}